# Developing a Vehicle Re-routing Algorithm using Connected Vehicle (CV) Technology

**Dr Mao Ye**
Research Associate
Department of Computer Science
Loughborough University, Loughborough, UK
Email: m.ye@lboro.ac.uk

**Dr Nicolette Formosa, corresponding author**
Research Fellow
Transport and Urban Planning
School of Architecture, Building and Civil Engineering,
Loughborough University, Loughborough, UK
Email: n.formosa@lboro.ac.uk

**Professor Mohammed Quddus**
Chair in Intelligent Transport Systems
Centre for Transport Studies
Department of Civil and Environmental Engineering
Imperial College London, London, UK
Email: m.quddus@imperial.ac.uk

**ABSTRACT**
Vehicle Ad-hoc Networks (VANETs) act as the core of vehicular communications and provide the fundamental wireless communication architecture to support both vehicle-to-vehicle (V2V) and vehicle-to-infrastructure (V2I) communication. Therefore, by leveraging only communication technologies, Connected Vehicles (CVs) can navigate through the dynamic road network. However, such vehicles are still in their infancy but are expected to have a significant impact on safety and mobility such as reducing non-recurrent congestion in case of a vehicle breakdown or other roadway incidents. To evaluate their impacts, this research examines the benefits of having CVs when a vehicle breakdown occurs by developing an intelligent proactive re-routing algorithm. Due to a lack of real-world data, this paper adopts an integrated simulated framework consisting of a V2X (OMNET++) communication simulator and a traffic microscopic simulator (SUMO). The developed algorithm functions such that when a vehicle is broken down within a live traffic lane, the system detects the breakdown, generates warning messages immediately and transmits them to approaching vehicles. Based on the real-time notification, informed vehicles proactively re-route to alternative roads to avoid the breakdown zone. Two scenarios were developed where a breakdown occurs within and outside a junction for both V2X-enabled and disabled systems. Results show that V2X-enabled CV re-routing mechanism can improve the traffic efficiency by reducing congestion and enhance the traffic safety by smoothing accelerations and decelerations of affected vehicles with low infrastructure costs. The algorithm would be useful to highways agencies (Department for Transport) and vehicle manufacturers in introducing CVs onto existing road networks.

## INTRODUCTION

Connected vehicles (CVs) can enhance the information gathered from their own sensors, with information from surrounding vehicles and dynamic environments for decision making through wireless communication technologies. Such technologies have the potential to improve traffic efficiency by identifying traffic congestion in a timely manner. The basic principle to alleviate congestion is that CVs share essential data and information about the current local traffic situation and use this information to optimise their routes. Nevertheless, while recurrent congestion can occur repeatedly due to similar patterns (i.e., rush hours), non-recurrent congestion originates from unexpected events such as incidents or vehicle breakdowns (*1*). Different from recurrent congestion which can be predicted, non-recurrent congestion is much challenging with more uncertainty. Real-time traffic flow detection and information sharing can be potential solutions for handing the non-recurrent congestion.

England's strategic road network (SRN) encounters non-recurrent congestion (one third of the total (*2*)) due to unexpected vehicle breakdowns, which pose a significant negative impact on traffic efficiency and safety. Especially, traffic flow approaching the motorway junctions and road works with changeable movements and high congestion probably falls into some unexpected events. By employing some wired technologies such as inductive loop detectors (ILDs), probe vehicles, radar and video cameras, the dedicated traffic management systems can continuously monitor and detect the abnormal status of road traffic. Since the installation and maintenance cost of wired technologies are quite high, vehicle-to-everything (V2X) technologies would be a cheaper and faster alternative for road traffic management and hazard warning. In V2X, each vehicle and roadside units (RSUs) acting as nodes can exchange essential data (e.g., vehicle type, position, speed, and acceleration) and warning messages (e.g., hazardous weather information, sudden hard breaking notification, traffic signal violation warning and closed road warning) among them by uniform or heterogeneous wireless communications (*3*). Such data can be transmitted in real time (with delay of less than one millisecond) to nearby RSUs and traffic control centres (TCCs).

Vehicle breakdowns within the SRN are responsible for causing significant delays and collisions. It is also more detrimental if a breakdown occurs within an approach area of roadworks or a slip road. Detecting any breakdowns as quickly as possible facilitates the design of traffic management interventions aimed at reducing traffic congestion and enhancing safety. One of the potential technologies is Vehicular Ad hoc Networks (VANETs). VANETs can efficiently bridge intelligent traffic system (ITS) and internet of vehicles (IoVs) to depict the future blueprint of road traffic. VANETs build a distributed inter-vehicular communication network, which is autonomous, cost efficient and based on flexible architectural design principles (*4*). Therefore, by enabling vehicular communications, the warning messages of any vehicle breakdown can be easily disseminated to surrounding vehicles to avoid potential vehicle collisions and traffic congestion. In VANETs, vehicles can convey information to: (1) other vehicles using V2V communications through which a vehicle communicates directly with another vehicle and shares information related to traffic conditions, such as traffic jams or accidents, (2) RSUs that are connected to the Internet and can communicate with other RSUs and roaming vehicles, which transmit data using V2I communications. However, when RSUs are not directly in the range of vehicles, V2V communication is considered mandatory in VANETs.

Since testing new applications directly in real environments is typically not possible due to major infrastructure upgrades and a significant commitment of support staff, simulators play a vital role in the evaluation of the developed applications that use VANETs. In order to assess the efficiency of IVC on vehicle routing strategies, congestion resolution, safety measures, travel experience, and to optimise the solutions, both vehicular traffic simulations and network layer simulations should be concurrently performed. Vehicle traffic simulators generate the necessary, realistic demand for vehicular mobility, which is then used as an input in network simulators. On the other hand, network simulators represent vehicles as nodes in a wireless network to simulate data transfer to mimic data dispersion topologies between nodes. Due to the differences between traffic simulators and network simulators, a meditation framework is necessary to facilitate seamless operations between the two simulators.

As a result, due to a lack of real-world data, this paper made use of a simulation network. For this study, SUMO (*5*) was selected as the traffic simulator, OMNeT++ (*6*) as the network simulator and Veins toolkit (*7*) as the mediation framework between SUMO and OMNetT++. The primary objective of this paper is to examine whether vehicle-to-vehicle (V2V) and vehicle-to-infrastructure (V2I) communication technologies can be applied to targeted and event-driven traffic management systems. Therefore, an intelligent re-routing solution for CVs when accidents or roadblocks occur has been developed. In the simulated network, vehicle breakdowns were randomly created. Once the vehicle breakdown occurred, the warning messages were generated immediately and then transmitted to the approaching vehicles. Based on the real-time notification, informed vehicles proactively re-routed to alternative roads to avoid the breakdown site/ zone. The experiment was configured to investigate whether the V2X-enabled traffic management system could enhance traffic efficiency and traffic safety. The delays and deceleration rates were selected as the key performance indicators to evaluate the abovementioned V2X-based traffic management system at roadworks.

## LITERATURE REVIEW

VANET is becoming increasingly crucial in the current intelligent transport systems (ITS) with a mass of attention from academia and industry (*8*). VANET consists of a plethora of vehicles with wireless transceivers to communicate with each other information (such as vehicle trajectories, kinematic data, traffic light pattern) which be obtained quickly via V2V and V2I communications. The shared traffic information can be further incorporated with varying techniques to support traffic management (e.g., traffic congestion detection and traffic states prediction). Therefore, VANETs have the potential to improve traffic safety, enhance the driving experience and benefit the environment. To highlight their importance in literature, this section (1) investigates the simulation platforms available supporting VANETs-enabled applications and (2) reviews the approaches of using VANETs and re-routing solutions in advanced technologies to solve traffic-related issues.

### Simulation Platforms

Due to the high cost and high intensity of Field Operational Testing (FOT) when deploying VANET scenarios, virtual testing (simulation) is an alternative method to test the proposed models and protocols. The majority of the VANET studies and projects in the literature rely on experimental simulations to test their performance. Two core components should be considered when conducting simulations for VANET scenarios, (i) vehicular mobility (road traffic simulation) and (ii) vehicular communications (network simulation).

Realistic road traffic mobility is essential in VANET scenarios to ensure accurate and representative results. This is carried out in vehicular traffic simulator which equips the network with real-time data collected by inductive loop detectors. By calibrating the simulation parameters, the derived results are expected to be as close as the observations of a real-world testing. To enable vehicular communication, a federated framework of traffic simulation and network simulation is essential. The traffic simulator is responsible for generating trace files for moving vehicles, and the network simulator can connect vehicles by wireless communications. The coupling level between traffic and network simulators is important to consider because it determines the way of data sharing. It can be categorized into three types, such as one-way decoupled, two-way (bi-directional) loosely coupled, and two-way (bi-directional) tightly coupled (*9*).

Veins framework is the latest state-of-the-art simulation platform, which federates all the benefits of both simulation tools, i.e., traffic and network simulation. Veins framework which is bi-directionally coupled combines SUMO and the INET project in OMNET++ via TCP connections (*7*). Veins can also support numerous traffic efficiency applications (e.g., intelligent traffic management) and traffic safety applications (e.g., emergency warnings). The integration in Veins between both simulators is bi-directional loosely coupled simulators. The majority of the VANET applications are simulated with this type of coupling because it is flexible and gives independence to the development of both simulators. Other simulators which adopt this approach include TraNS (*10*) and iTETRIS (*11*). However, there are other simulation platforms such as NCTUns which is a completely integrated simulator bringing both traffic and network simulation into a single tool (*12*).

**Traffic-related Approaches and Re-routing Solutions**

A number of advanced traffic management solutions have been proposed for traffic detection and estimation in the past years. In the literature, the traditional traffic management approaches are being replaced by ITS and VANETs.

Liu et al. (*13*), developed an algorithm which detects the real-time traffic flow and traffic congestion by using video cameras to count vehicles. Another developed method for real-time traffic estimation utilises the travel time of the vehicles via inductive loop detectors (*14*). To distinguish the levels of traffic congestion, a fuzzy-logic-based mechanism was also proposed (*15*). Similarly, a control algorithm based on fuzzy logic, which collect and utilise the information of each vehicle through V2I, was proposed in AUTOPIA system to reduce traffic congestion (*16*). Other studies used beacon messages to estimate the traffic conditions. For example, Xu et al. (*17*), estimatedthe traffic conditions by processing and interpolating bus travel time from one RSU to another through beacon messages sharing. CoTEC (a cooperative vehicle system) receives all the information from surrounding vehicles and adopts fuzzy logic to measure local traffic conditions (*11*). UCONDES is another proposal based on V2V, but uses an artificial neural network to detect and classify the congestion level (*18*). DIVERT was developed for congestion avoidance by a distributed vehicular traffic re-routing system (*19*). In DIVERT, a large part of the re-routing computation is offloaded to vehicles to balance the user privacy with the re-routing effectiveness.

In summary, current traffic detection methods make use of wired technologies such as loop detectors, video, camera, microphone, probe car, and if re-routing is required (e.g., in case of an emergency) this is generally communicated via a variable message sign. However, as the CV technology increases and CVs become more available, the information sharing via vehicular communications can transform the way of traffic status detection and re-routing. One new approach is by adopting a V2X-enabled system. Current literature lacks such investigation. Therefore, to appraise the benefits of this approach, a vehicle re-routing algorithm using CVs was developed in a unified simulation platform (Veins). Veins allows various road scenarios to be implemented using real-world data. Moreover, most of the CVs-enabled rerouting scenarios developed in the literature are test grid-based or urban road layouts. This research made use of a realistic highway scenarios with complex roadworks occurring to implement and test the developed rerouting mechanism.

## METHODOLOGY

In this section, an enhanced methodology for the vehicle re-routing aided by V2V technologies is described. This is then followed by a demonstration of a research scenario and the corresponding experimental framework are specified. Additionally, the inputs and detailed parameters are also discussed.

The CV re-routing mechanism is possible due to exchange of data and information supported by the V2V technology. By sharing period messages based on V2V technology among adjacent vehicles, each vehicle can easily monitor the trajectories of oncoming vehicles and achieve some safety-related functions in real time. Thereafter, with the embedded decision-making module, each vehicle executes the re-routing based on the shared information.

To achieve a V2X enabled traffic management system, one unified experiment platform known as Veins was employed in our research. Veins is an excellent simulation platform which integrate two well-established simulators namely SUMO (a traffic simulator) and OMNET++ (a communication simulator). IEEE 802.11p and IEEE 1609.4 are applied in Veins to define the details of V2X communications. Compared to other traffic microsimulation software, real-time simulations can be performed online in Veins with two bi-directionally coupled simulators. Additionally, Veins can support numerous traffic efficiency applications (i.e., ITS applications) and traffic safety applications (i.e., emergency warnings). In our research, vehicular movements and traffic flow were configured and managed in SUMO. Veins and OMNET++ are primarily for network environment configuration, data collection, data processing, accident generation and vehicle rerouting. Both SUMO and OMNET++ simulators have been extended by modules allowing the road traffic simulation to link with its network simulation counterpart via a TCP/IP connection (*20*). This simulator extension allows the network simulation for vehicular communications to directly control the road traffic simulation in real time and investigate the influence of V2X-enabled solutions. Veins, as a fine-grained control interface, bridges the SUMO and OMNET++ for real-time applications. The unified simulation platform is schematically shown in **Figure 1**.

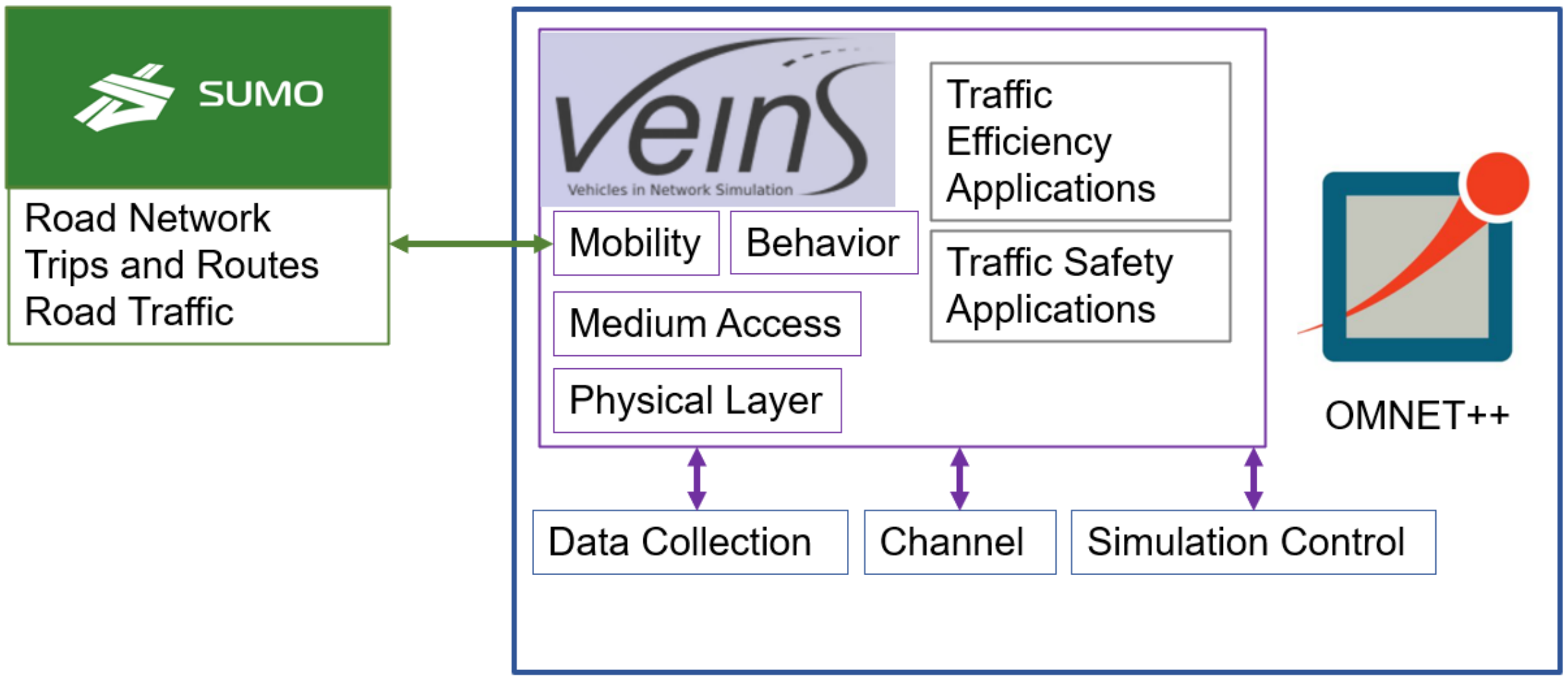

**Figure 1 Configuration of V2X-enabled traffic management system**

Within the whole simulation framework, several assumptions and conditions need to be considered. It is assumed that all vehicles in the simulation have the on-board unit for data processing and networking, and they can obtain their own kinematic information from a digital map and a GPS receiver. The original trajectory of each vehicle is predefined before the simulation execution. Each vehicle is assumed to have the same communication range, which is pre-defined in the OMNET++ simulator. During the process of beacon broadcasting, we assume that vehicle movements are paused at that simulation step and the vehicle restarts to move between the interval of two broadcasts. Each vehicle can periodically transmit the packets without any error. The vehicle movement is assumed to obey and follow a predefined mobility model with the parameters shown in **Table 1** in next section. In addition, the assumptions of specific settings are described in **Table 1** and **Table 2**.

Based on the integrated simulation platform and assumptions, an event-driven traffic algorithm for CV re-routing was developed. It starts functioning once a vehicle breakdown occurs which causes the broadcasting mechanism to awaken. Warning messages are then generated immediately and then transmitted to the neighbouring vehicles in the form of periodical beacons. Two main pseudocodes of the developed CV-based re-routing algorithm and corresponding descriptions are presented in **Figure 2** and **Figure 3**. The pseudo-code for the vehicle breakdown detection and warning messages dissemination is given in **Figure 2.**

**Algorithm 1: Vehicle breakdown detection and warning messages dissemination**

```
initialize (){
if (breakdownCount  > 0){
    simulationTime breakdownStart = Breakdown Start;
    startBreakdownMsg = Message ("Breakdown");
    stopBreakdownMsg = Message("BreakdownResolved");
    scheduleAt(simulation_Time() + breakdownStart, startBreakdownMsg);
        }
}

handleWarningMsg(){
if (msg == startBreakdownMsg){
    setSpeed(0);
    simulation_time breakdownDuration = Breakdown Duration;
    scheduleAt(simulation_Time() + breakdownDuration, stopBreakdownMsg);
    breakdownCount--;
}
else if (msg == stopBreakdownMsg){
    setSpeed(-1);
        if (breakdownCount > 0){
                simulation_time breakdownInterval = Breakdown Interval;
                scheduleAt(simulation_Time() + breakdownInterval, startBreakdownMsg);
        }
}
```

**Figure 2 Pseudo-code for vehicle breakdown detection and warning messages dissemination**

The breakdown warning messages are continuously and periodically broadcast until the clearance of the vehicle breakdown. If the approaching vehicles receive the warning messages via

V2V, these vehicles firstly relay the warning messages to the neighbouring vehicles and then proactively trigger the dynamic re-routing mechanism based on the Dijkstra's algorithm to avoid the vehicle breakdown site. The new route would be re-computed by Dijkstra's algorithm with updated parameters and the same destination. If there are no alternative paths, approaching vehicles travel with caution to avoid any collisions. The pseudo-code for the warning message receiving and proactive CV rerouting is displayed in **Figure 3.**

**Algorithm 2: Warning messages receiving and proactive CV rerouting**

```
receiveWarningMessage();
getCurrentRouteId();
getCurreentPositionOnRoute();
getRouteEdgeIds();
getBreakdownPosition();
changeRoute(){
 if (travelTime >= 0){
        newTime = traveltime;
        setVehicleVariable(newTime, edgeId, variableType, nodeId, variableId);
        reRoute();
 }
else{
        keepVehicleVariable(edgeId, variableType, nodeId, variableId);
    }
changeVehiclePosition();
```

**Figure 3 Pseudo-code for Warning messages receiving and proactive CV rerouting**

A summary of the event-driven traffic management system for CV re-routing is presented in **Figure 4**. The experiments were configured to investigate whether the V2X-enabled CV rerouting mechanism could enhance traffic efficiency and traffic safety. To analyse and evaluate the CV re-routing performance, traffic delays and deceleration rates were selected as the key performance indicators in this research.

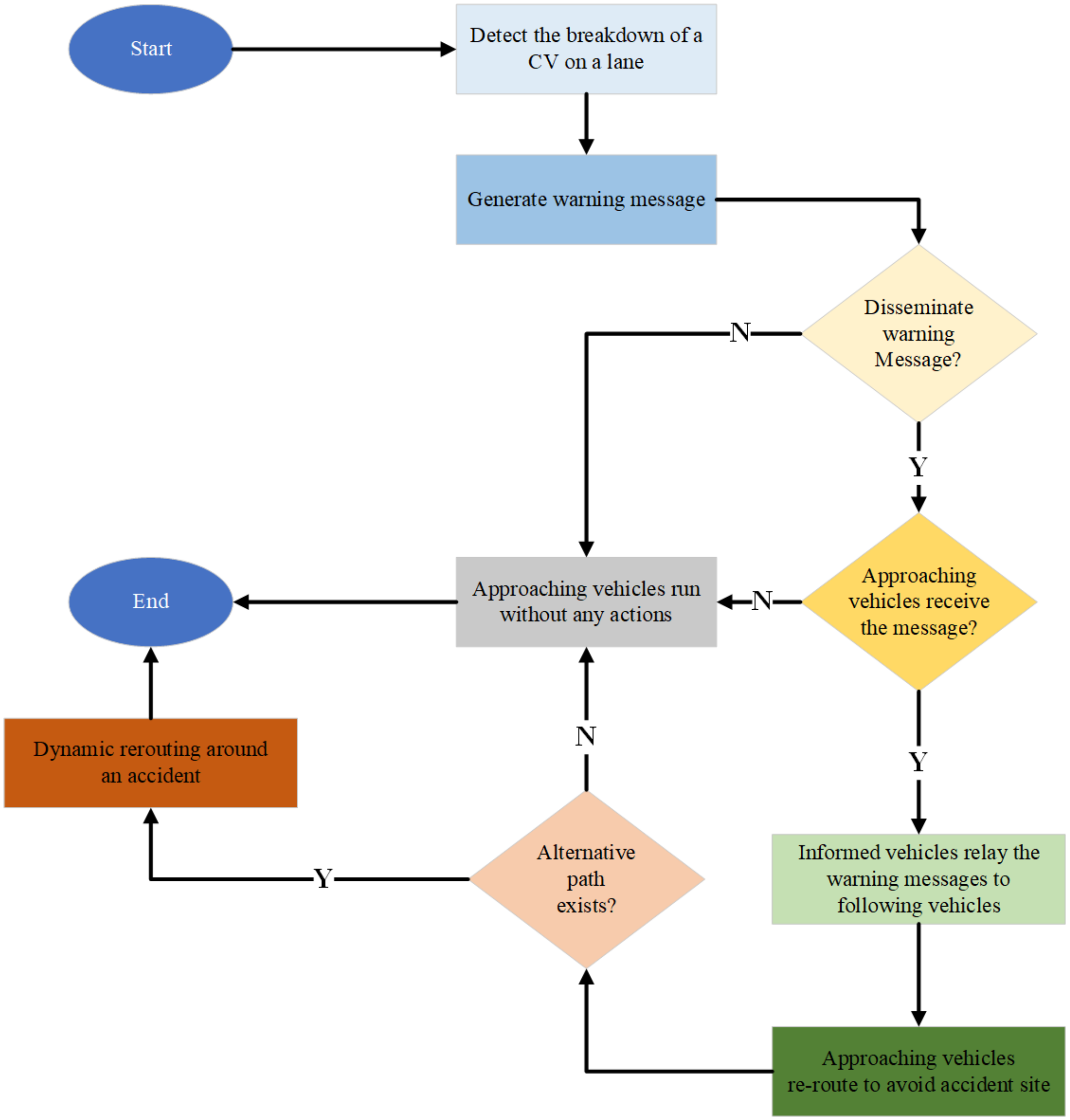

**Figure 4 Flowchart of the event-driven traffic management system**

## DATA AND VEHICLE BREAKDOWN SCENARIOS

Speed variations approaching the highway junction and roadworks might incur unexpected accidents or breakdown of vehicles. These two factors were considered to select the research scenarios. To evaluate the developed CV re-routing mechanism in and around junctions, one road segment (Junction 17 to Junction 16 of M1 motorway in the UK) leading to roadwork zone is simulated and analysed in this research. Junction 13 - 16 is currently being converted to smart motorway, and the roadworks start about 3km from Junction 16 in the southbound direction. This part of the motorway segment is an ideal test site for simulating V2X-related scenarios. Therefore, a simulation model consisting of about 5km stretch leading to Junction 16 was created and calibrated in Veins. The simulation calibration is generally divided into two parts, road layout calibration and traffic demand calibration.

**Figure 5 (a)** presents the original road map extracted from OpenStreetMap data of the UK M1 motorway which was modified by the visual network editor NetEdit. To make the experimental scenario concise and accurate, irrelevant roads and regions in the original map were deleted. Moreover, the layers of the overpass around the junctions were also amended. The calibrated road layout in SUMO is shown in **Figure 5 (b)**.

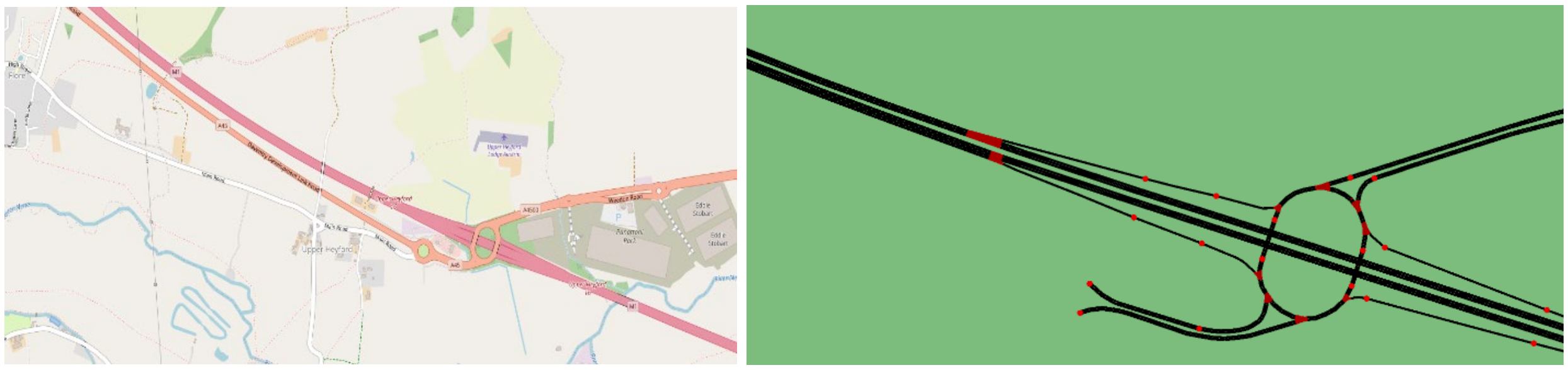

**(a)** **(b)**

**Figure 5 Southbound road segment leading to M1 Junction 16 from (a) Open Street Map and (b) SUMO**

With regards to the traffic demand, one hour peak-time (4pm-5pm) traffic flow nearby the selected motorway segment was chosen to reflect the simulation configuration in this research. The real-time traffic data was obtained from National Highways, and the original data was further calibrated by the field test in this research. An instrumented vehicle equipped with multiple sensors, as the probe vehicle, was driven on the test route to collect kinematic data. Vehicle-based speed profiles and journey times were employed to verify the traffic model in the simulation. Based on the official data and field test, the calibrated traffic information is presented in **Table 1**. From **Table 1**, MinGap describes the offset to the leading vehicle during a congested traffic condition and sigma, the parameter in car-following model, indicates the driver imperfection (0 denotes 'perfect driving'). Normally, the drivers of HGVs are better trained than ones of passenger vehicles.

**TABLE 1 Calibrated traffic information in SUMO**

| **Traffic demand parameters** | **Value** |
|---|---|
| Total number of vehicles | 400 |
| The percentage of passenger vehicles | 80% |
| The percentage of HGVs | 20% |
| Acceleration | 2.6 $m/s^2$ |
| Deceleration | 4.5 $m/s^2$ |
| Max speed | 60 mph |
| MinGap | 2.5m |
| Sigma | 0.6 for passenger vehicles<br>0.4 for HGVs |
| Car-following model | Krauss |
| Lane-Changing model | LC2013 |
| Simulation time | 1000s |

After calibrating and validation of the setting of traffic model, the network configurations for vehicular communications were enabled. The relevant parameters for communication setting are described in **Table 2.**

**TABLE 2 Network configuration in OMNET++**

| Network parameters | Value |
|---|---|
| Beacon interval | 1s |
| Propagation model | Two-rays ground reflection model |
| Transmission range | 300m |
| Transmission power | 20mW |
| Antenna height | 1.895m |
| MAC layer protocol | IEEE 802.11p DSRC |
| Packet size | 1KB |
| Simulation time | 1000s |

Two scenarios were developed in the simulation model to measure the influence of vehicular breakdown in and round the junctions. Scenario 1 describes a vehicle breakdown occurred inside Junction 16 (see position A in **Figure 6**) and scenario 2 is made up of a vehicle breakdown outside Junction 16 (see position B in **Figure 6**). Data from inductive loop detectors (ILDs) located in position C (in **Figure 6)** was also adopted to measure the traffic flow.

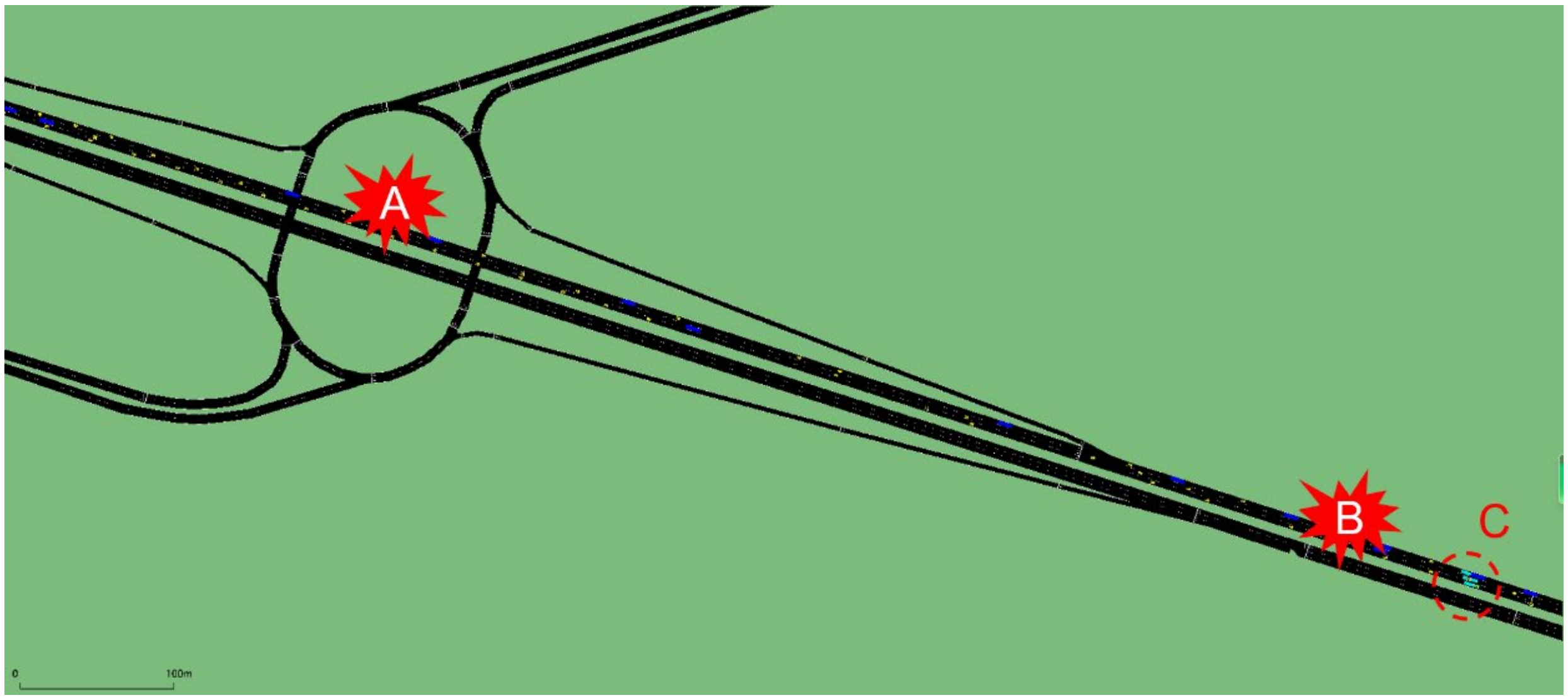

**Figure 6 Two demonstrated scenarios with embedded ILDs**

For both scenario 1 and 2, three individual experiments were conducted to explore the benefit of CAV rerouting mechanism. **Table 3** provides the detailed description of each experiment. Each experiment has the same layout and configuration. Experiment 1 and 4 are the baseline experiments for Scenario 1 and 2 respectively.

**TABLE 3 Comparative experiments for two scenarios**

| | | |
|---|---|---|
| **Scenario 1** (breakdown within junction) | Experiment 1 | Normal traffic without vehicle breakdown |
| | Experiment 2 | Vehicle breakdown with V2X-disabled system |
| | Experiment 3 | Vehicle breakdown with V2X-enabled system |
| **Scenario 2** (breakdown outside junction) | Experiment 4 | Normal traffic without vehicle breakdown |
| | Experiment 5 | Vehicle breakdown with V2X-disabled system |
| | Experiment 6 | Vehicle breakdown with V2X-enabled system |

## RESULTS

Each experiment presented in **Table 3** was tested in the integrated simulation framework developed in this study. This section provides the key findings from the experiments when compared to baseline scenarios. The scenarios were tested to evaluate the introduction of CV technology with respect to: (1) potential benefits of adopting V2X rerouting systems, (2) how these systems are beneficial for traffic efficiency in terms of delays and congestion, and (3) if they are able to enhance safety.

### Benefits of V2X-enabled CV re-routing mechanism

The usage of V2X-enabled CV rerouting mechanism can efficiently prevent the traffic congestion and improve the traffic safety with low infrastructure costs. This type of rerouting mechanism trends to be more practical and popular with the high market penetration rate (MPR) of CVs in the future. To highlight these benefits, the simulation framework was adopted to compare Experiment 3 with Experiment 2 in Scenario 1 and are presented in **Figure 7**.

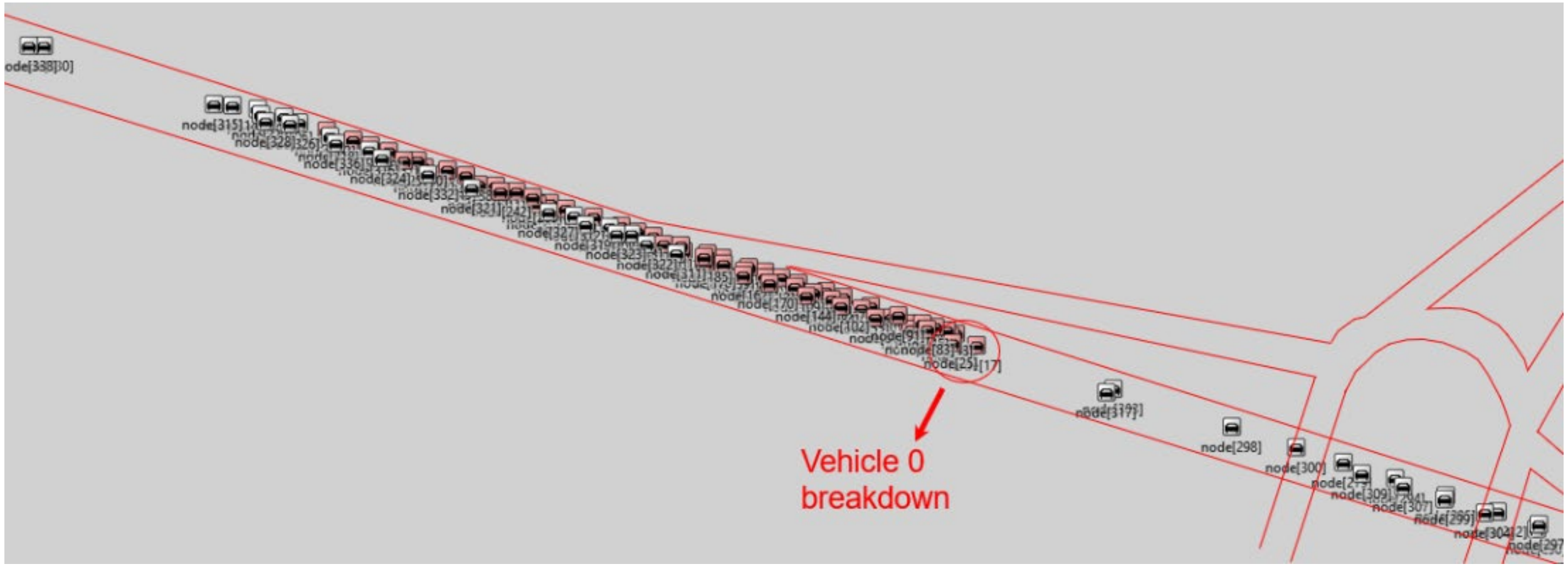

(a) Vehicle 0 breakdown in Experiment 2: the queue built up following the breakdown

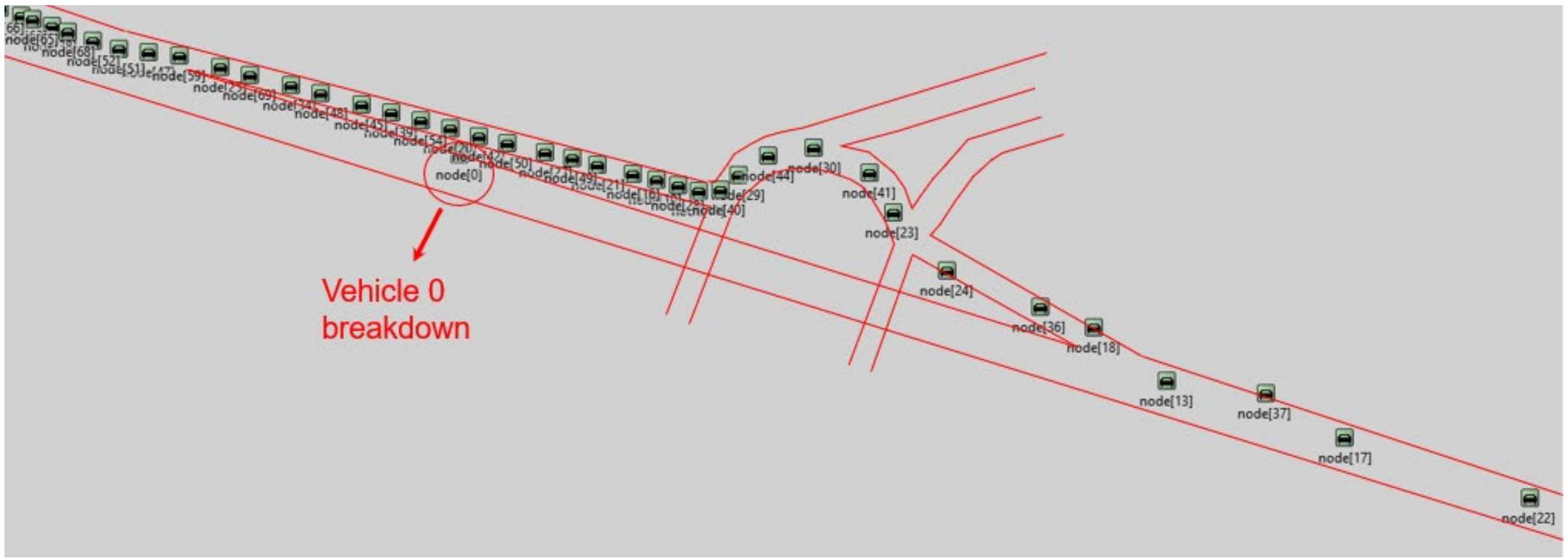

(b) Vehicle 0 breakdown in Experiment 3: CVs diverted to the alternative route once they received the warning messages

**Figure 7 Comparisons of traffic status in Experiment 2 and 3**

By comparing the findings from each experiment shown in **Figure 7**, it can be easily deduced that V2X-enabled traffic management system could proactively reroute CVs to move around the vehicle breakdown site. However, without V2X communications, the network resulted in a large queue of stopped vehicles highlighted in red (see **Figure 7 (a)**). This congestion blocked the road leading to the breakdown site and caused an increase in delays and journey times.

**Comparison of vehicle delays in Scenario 1 and 2**

Delays can lead to an increase in journey times. In order to calculate delays caused by a vehicle breakdown in both Scenario 1 and 2 (within and outside junction respectively), a normal traffic without a vehicle breakdown were selected as the baseline simulations (i.e., experiments 1 and 4). For vehicle breakdowns within junctions, (Scenario 1) the vehicle delays as presented in **Figure 8**. The V2X-enabled CAV rerouting mechanism in Experiment 3 is superior to V2X-disable one in Experiment 2. In fact, average delays in Experiment 2 are 248s, which is about 8.5 times higher than that of Experiment 3 with 29s. This highlights the importance of adopting a V2X rerouting system not only with regards to reducing journey times, but this also affects the environment, resulting in a reduction in carbon emissions and a better air quality.

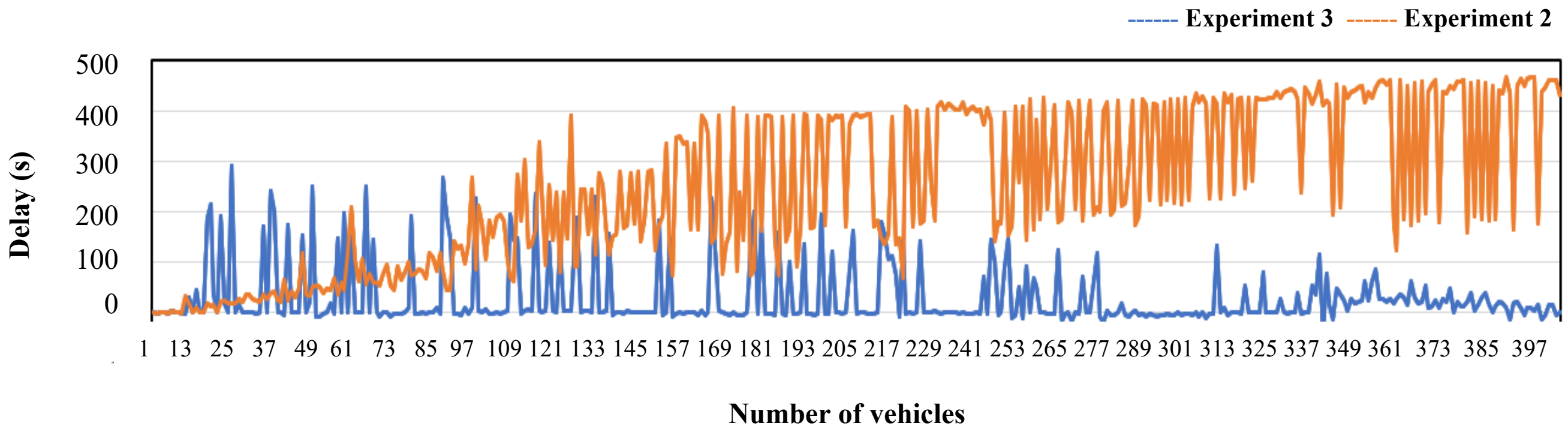

**Figure 8 Comparison of time delay in Experiment 2 and 3**

The vehicle breakdown site in Scenario 2 is outside the Junction16. Experiments 5 and 6 were compared, and the corresponding total journey time of 400 vehicles are presented in **Figure 9**. Results show that with the V2X-enbaled CAV rerouting mechanism (Experiment 6) the total journey time averagely decreases by about 10s. This shows that by employing the V2X communications the approaching vehicle were able to find alternative routes to avoid the breakdown site and were able to change lanes early to ease traffic congestion.

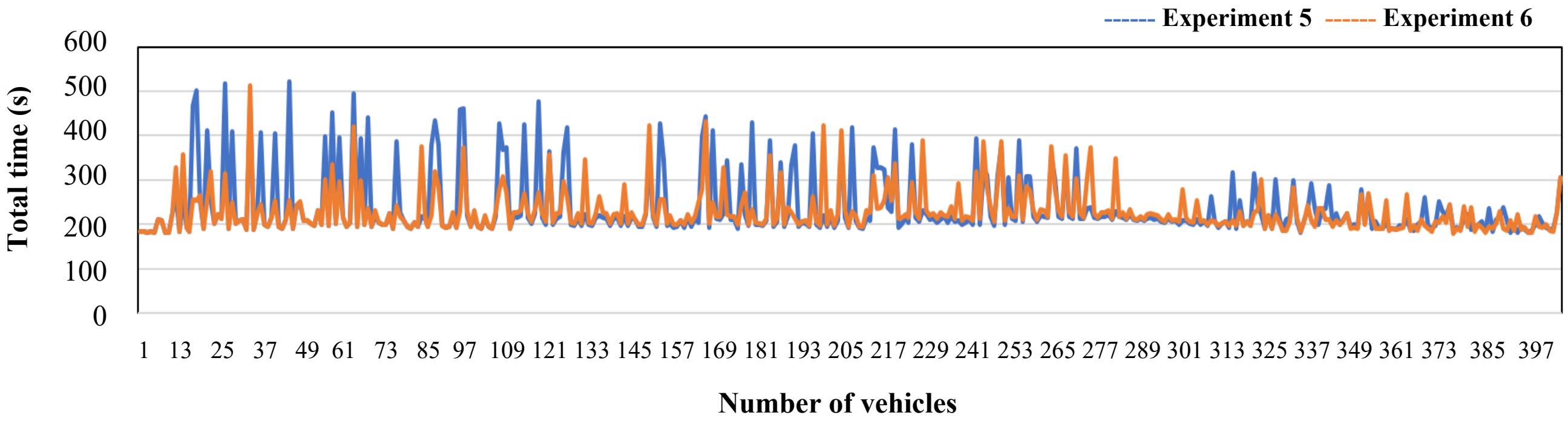

**Figure 9 Comparison of total time in Experiment 5 and 6**

**Comparison of speed and acceleration profiles of vehicles approaching breakdown site**

Multiple factors affect road traffic efficiency and safety. Some of the influential factors which have a major impact are speed and acceleration. Nevertheless, different combinations of speeds and accelerations result in a varying traffic status. For example, when a vehicle breaks down, the network may become highly congested where the surrounding vehicles are travelling with low speed and variable acceleration. The effect of a broken-down vehicle (Vehicle 0) on speed and acceleration was also studied on the surrounding vehicles in the simulation as shown in **Figure 10**.

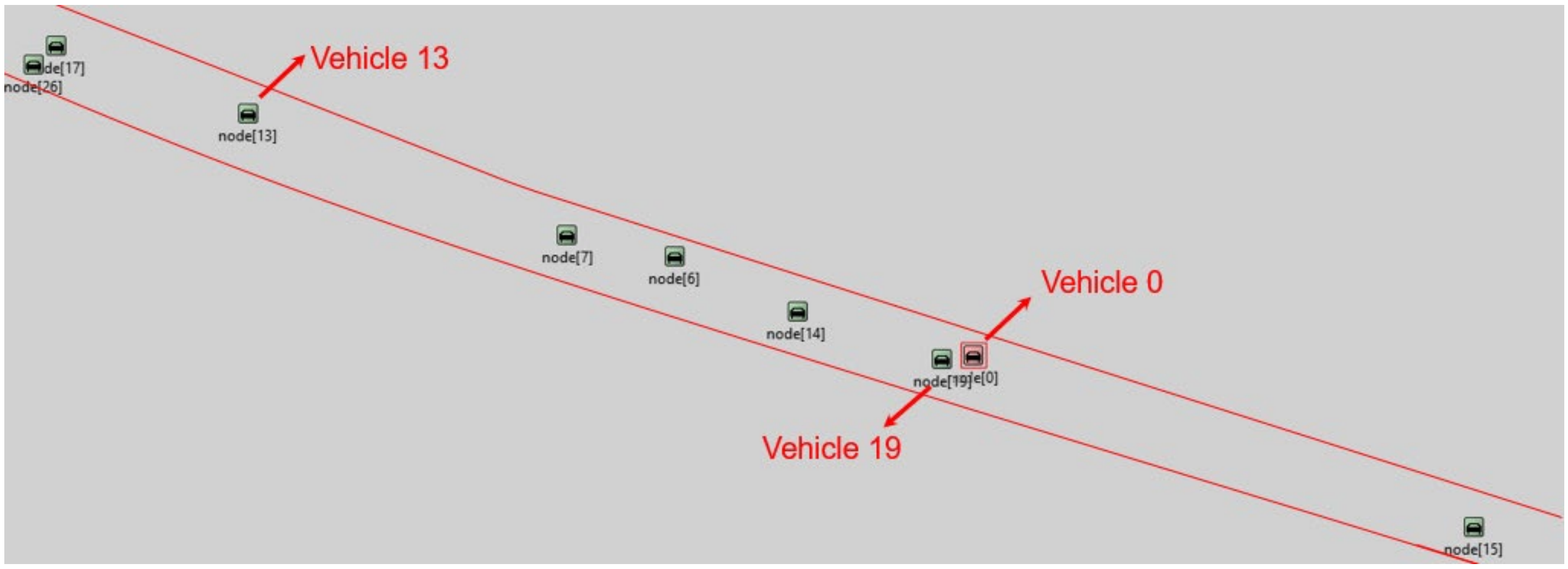

**Figure 10 Breakdown - Vehicle 0**

From **Figure 10** when Vehicle 0 broke down in Experiment 2 and 3 Vehicle 19 was very close to the breakdown vehicle. From the acceleration and speed profile presented in **Figure 11(a)** and **(b)** respectively, Vehicle 19 did not have enough time to react to the breakdown (at epoch 115). This occurred even though Vehicle 19 received the message from Vehicle 0. In comparison, Vehicle 13, which was further away from breakdown site had enough time to react to the message received from Vehicle 0. Its acceleration and speed profiles can be observed in . This is also reflected when comparing the average acceleration of both Vehicle 13 and 19, 0.540 $m/s^2$ and 0.448 $m/s^2$ and corresponding variances 0.552 and 0.405 respectively.

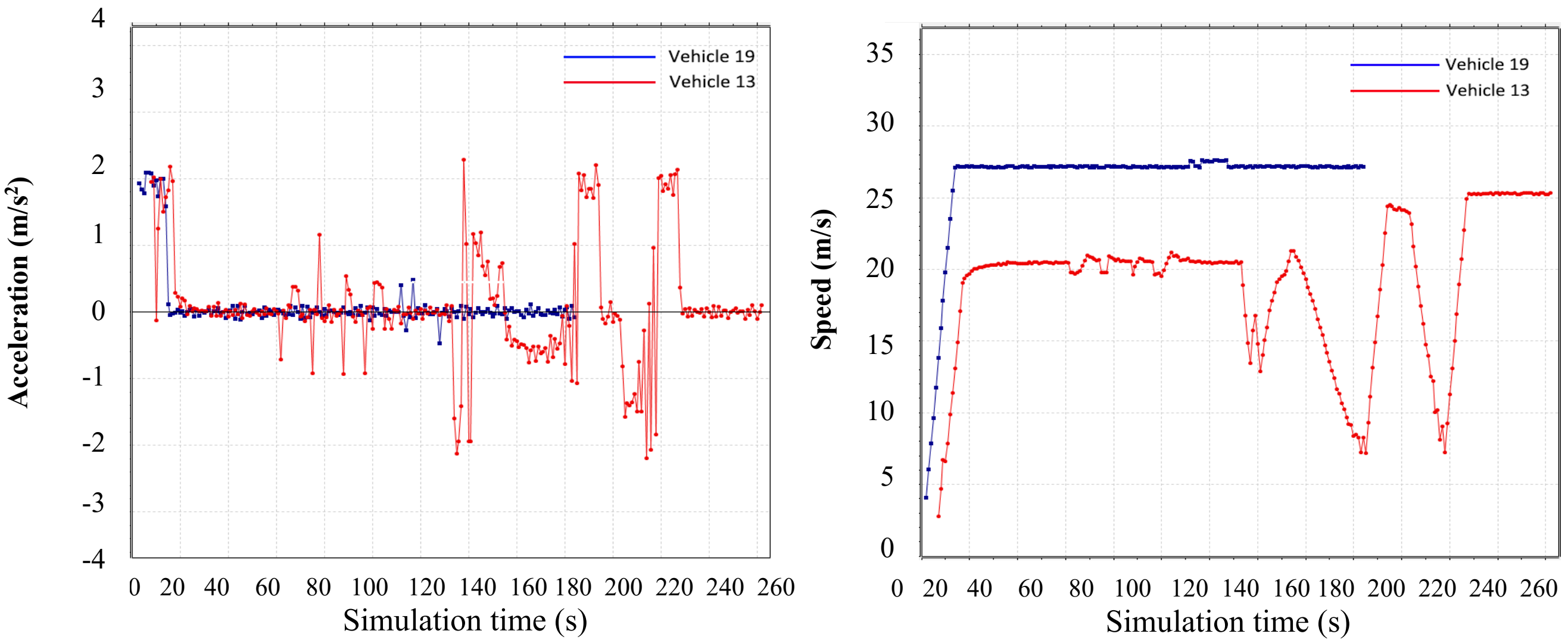

**Figure 11 Comparison of (a) acceleration and (b) speed profile of Vehicle 13 and 19**

**Analysis of braking behaviours**

An influential factor used to identify traffic conflicts is the deceleration rate (*21*). From the simulation, it was possible to observe how the declaration rate varies by scenario. It was found that the average deceleration rate in Scenario 1 (i.e., breakdown within junction) was significantly lower than that of Scenario 2 (i.e., breakdown outside junction). This is because approaching vehicles were able to find alternative routes to avoid the breakdown site in Scenario 1. The detailed analysis of deceleration rates of three experiments in Scenario 1 is shown in **Table 4**.

When comparing the baseline scenario (Experiment 1) with a V2X-disbaled system (Experiment 2) for Scenario 1, the average deceleration and variance surge, by 231.78% from 0.107 $m/s^2$ to 0.355 $m/s^2$, by 42.17% from 0.230 to 0.327, respectively. On the other hand, when comparing the baseline scenario (Experiment 1) with V2X-enabled system (Experiment 3) the CAV rerouting was found to increase the average deceleration rate by 118.69% (from 0.107 $m/s^2$ to 0.234 $m/s^2$) and the variances of vehicles' deceleration rates increases by 3.91% (from 0.23 to 0.239). While there is an increase in both cases due to the traffic interactions at the breakdown site, when comparing between Experiment 2 and 3 it shows that increase in deceleration rate is approximately only a half when V2X is enabled, and the variation of speed is relatively significantly smoother. This is essential to enhance safety on the road network.

**TABLE 4 Comparison of deceleration rates in Scenario 1**

| Deceleration | Mean ($m/s^2$) | Percentage change | Variation in speed | Percentage change |
|---|---|---|---|---|
| **Experiment 1** | 0.107 | | 0.230 | |
| **Experiment 2** | 0.355 | 231.78% | 0.327 | 42.17% |
| **Experiment 3** | 0.234 | 118.69% | 0.239 | 3.91% |

In summary, it can be concluded that V2X-enabled CAV re-routing mechanism has a great potential in enhancing safety and easing traffic congestion following vehicle breakdowns both within and outside junction. This is because it aims to optimise the vehicles' trajectories instantaneously by exchanging positional and vehicle kinematics.

## DISCUSSION AND CONCLUSIONS

This research developed a CV re-routing algorithm to test representative scenarios such as vehicle breakdowns within and outside a junction. The algorithm was developed in an integrated simulation platform consisting of  a V2X simulator and a traffic micro-simulator. This included the simulation of communication between vehicles and infrastructure to facilitate cooperative actions within traffic environments. This integration required elements to be combined from SUMO, which represents a microscopic traffic simulator, with OMNet++ which allows for the distribution of packets of information between vehicles and infrastructure (e.g., roadside units).

A simulation model representing a motorway corridor (M1 Motorway Junction 17 to Junction 16) was created within the integrated simulation platform to evaluate the implications of vehicle connectivity during stopped vehicle incidents (e.g., car breakdowns) resulting in temporary lane closures. The automated CV re-routing solution works as an event-triggered algorithm which notifies adjacent vehicles about emergent events and provides alternative routes to enhance traffic safety and avoid traffic congestion. Each vehicle was assumed to equip with a GPS receiver and wireless communication devices offering the opportunity for CVs to receive information regarding the lane closure and plan accordingly (e.g., detouring or changing lane). The impacts on delays and vehicle speeds as compared to situations when no vehicle connectivity was present, were evaluated to determine the benefit of its rerouting capabilities. Due to the event-triggered feature, this algorithm can further save wireless communication resources.

Results from the different experiments show that V2X-enabled CV rerouting mechanism improves traffic efficiency by reducing the delay and enhance traffic safety by smoothing the acceleration and deceleration rates. Apart from benefiting the CV rerouting, V2X-enabled messages are also practical in emergency warning to enlarge the sight of drivers and better the traffic safety. Drivers receiving the warning messages can also proactively react on the breakdown event even if there exists no alternative path. For example, efficient speed control and manoeuvre by drivers would be useful for avoiding potential collisions.

In addition to the risks presented by junctions themselves, additional complexity arises when a vehicle suddenly breaks down. Identifying a stopped vehicle quickly can prevent an imminent collision. The simulation model developed indicated that one of the potential technologies would be vehicle-to-vehicle (V2V) capability in which an 'imminent hazard warning' can be transmitted to nearby vehicles within 1 second. Furthermore, signals could also be transmitted to other roadside units (RSUs) such as variable message signs and regional traffic control centres. This will enhance both situational awareness and decision-making under uncertainty.

Although a decentralised approach (V2V) would be preferable as the market penetration rate (MPR) increases, CVs are still in their infancy. An SMMT projection in 2019 showed that CVs would account for about 30% of vehicle sales by 2035 (*22*). Since the objective of the decentralised approach aims to optimise vehicles' trajectories by exchanging positional and vehicle kinematics instantaneously, with the current CV MPR, the decentralised approach might not be the most feasible option to reflect the short-term reality but will be necessary as the MPR increases.

On a network with vehicles driving at high speeds, locating car breakdowns requires a methodical effort to locate, respond to, and prevent road accidents while re-establishing safe traffic capacity in a reasonable timeframe. This research fills the gap since it quantifies the impact of CVs in the event of a breakdown within and outside junctions while maximising traffic performance and highway safety through the development of a traffic simulation system. Results from this study will be beneficial to highways agencies as well as vehicle manufacturers (OEMs) and the Department for Transport to manage CV deployment.

**ACKNOWLEDGMENTS**

This paper is based on a project (CAVIAR) commissioned by National Highways (UK). The authors would like to express their special thanks to John Matthewson (from National Highways) and Jon de Souza and Eugenie Blyth (both from Galliford Try, the project partner) for their assistance in this research during the project. The opinions in this paper are those of the authors and do not necessarily reflect those of the National Highways or Galliford Try. The authors remain solely responsible for any errors or omissions.

**AUTHOR CONTRIBUTIONS**

The authors confirm contribution to the paper as follows: study conception and design: M. Ye, N. Formosa, M. Quddus; data collection: M. Ye, N. Formosa, M. Quddus; analysis and interpretation of results: M. Ye, N. Formosa, M. Quddus; draft manuscript preparation: M. Ye, N. Formosa, M. Quddus. All authors reviewed the results and approved the final version of the manuscript.